\documentclass[12pt,a4paper]{article}

\usepackage[utf8]{inputenc}
\usepackage[T1]{fontenc}
\usepackage{lmodern}
\usepackage{amsmath,amssymb}
\usepackage{cite}
\usepackage{hyperref}
\usepackage{graphicx}
\usepackage{caption}
\usepackage{subcaption}
\usepackage{float} 

\title{Benchmark Study of CEvNS Nuclear Recoil Observables for B, Mg, Ti, and Zr Targets Using Geant4} 

\author{Yusuf Havvat\\[2mm]
\small Department of Physics, Cukurova University, 01330 Adana, Turkey\\
\small \texttt{havvatyusuf249@gmail.com}
}

\date{} 

\begin{document}
\maketitle

\begin{abstract}
Coherent elastic neutrino--nucleus scattering (CEvNS), first predicted by Freedman~\cite{freedman1974} and experimentally observed by the COHERENT Collaboration~\cite{akimov2017,akimov2021}, provides a well-defined framework for studying nuclear recoil observables in low-energy neutrino interactions. 
In this work, we present a detector-agnostic Monte Carlo benchmark study of CEvNS-induced nuclear recoils for four representative target nuclei (B, Mg, Ti, and Zr), performed under fully identical and controlled simulation conditions using the \textsc{Geant4} toolkit~\cite{agostinelli2003,allison2016} and the \textsc{ROOT} analysis framework~\cite{brun1997}.

Rather than optimizing a specific experimental detector configuration, the analysis deliberately isolates intrinsic nuclear-physics effects by employing a simplified and transparent geometry. This approach enables a systematic comparison of recoil-energy spectra, nuclear form-factor suppression, and angular distributions on equal footing across a broad nuclear-mass range. Nuclear structure effects are consistently treated through the Helm form factor~\cite{helm1956}, allowing coherence loss and momentum-transfer dependence to be evaluated in a controlled manner.

The results highlight complementary target-dependent trends relevant for CEvNS detection strategies: lighter nuclei exhibit extended recoil-energy endpoints and reduced form-factor suppression, while heavier nuclei benefit from enhanced interaction rates driven by approximate $N^{2}$ scaling at low recoil energies. By decoupling detector-response effects from intrinsic CEvNS kinematics, this benchmark-level study establishes a physically transparent reference for target-material selection and provides guiding principles for future low-threshold CEvNS detector designs.
\end{abstract}

\section{Introduction}

Coherent elastic neutrino--nucleus scattering (CEvNS) was first predicted by Freedman in 1974~\cite{freedman1974} and experimentally observed more than four decades later by the COHERENT Collaboration using a pion decay-at-rest neutrino source at the Spallation Neutron Source (SNS)~\cite{akimov2017,akimov2021}. In CEvNS, a low-energy neutrino interacts coherently with an entire nucleus via a neutral-current process, leading to an enhanced cross section that approximately scales with the square of the neutron number, $N^2$. This property makes CEvNS a well-defined framework for studying nuclear recoil observables in low-energy neutrino interactions~\cite{scholberg2006,miranda2020}.

The experimental observation of CEvNS has motivated a variety of detector concepts based on different target materials, threshold technologies, and readout schemes. While most existing CEvNS experiments focus on a limited set of detector materials, comparative studies that isolate the impact of nuclear properties—such as target mass and form-factor suppression—remain useful for understanding general trends in CEvNS-induced recoil observables. In this context, simplified simulation studies can provide controlled benchmarks that complement more experiment-specific detector designs.

Modern Monte Carlo toolkits, such as Geant4~\cite{agostinelli2003,allison2006,allison2016}, together with analysis frameworks like ROOT~\cite{brun1997}, enable detailed simulations of neutrino--matter interactions under well-defined and reproducible conditions. When combined with standard nuclear form-factor parameterizations, these tools allow systematic comparisons between different target nuclei without introducing experiment-dependent complexities.

In this work, we present a comparative Monte Carlo benchmark study of CEvNS-induced nuclear recoils for four target nuclei: boron (B), magnesium (Mg), titanium (Ti), and zirconium (Zr). These nuclei span a broad range of atomic masses and neutron numbers, allowing the study of how nuclear mass and Helm form-factor suppression~\cite{helm1956,lewin1996} influence recoil-energy spectra and angular distributions. A simplified detector geometry is employed deliberately to ensure a transparent comparison between target materials, rather than to model a specific experimental setup. The goal of this study is to provide a reference-level comparison of CEvNS recoil observables under identical simulation conditions, suitable for benchmarking and material-selection considerations in CEvNS-related simulations.

\section{Simulation Framework}

All simulations were performed using the Geant4 Monte Carlo toolkit (version 11.2.0) and the ROOT framework for data analysis and visualization. The primary goal of the framework is to provide a controlled and reproducible environment for comparing the CEvNS response of four candidate target nuclei: boron (B), magnesium (Mg), titanium (Ti), and zirconium (Zr).

\subsection{Neutrino Source Model}

In this study we adopt a deliberately simplified neutrino-source configuration in order to perform a controlled, reproducible, and transparent comparison of CEvNS recoil observables across different target nuclei. The primary goal is therefore not the prediction of absolute event rates for a particular experiment, but the isolation of nuclear-physics effects (target mass, neutron number, and form-factor suppression) under identical simulation conditions.

For the baseline comparisons, a monoenergetic electron-neutrino ($\nu_e$) beam with $E_\nu=10$~MeV is used unless otherwise stated. This choice removes the dependence on a specific source spectrum (reactor, SNS, or solar neutrinos) and avoids folding the recoil distributions with experiment-dependent flux shapes. In this way, differences observed between targets can be attributed directly to nuclear properties rather than to spectral features of a chosen facility.

The choice of a 10~MeV neutrino energy is motivated by its relevance to typical recoil-energy scales explored in reactor and stopped-pion CEvNS experiments, where nuclear recoils lie predominantly in the sub-keV to few-keV range. This energy also avoids strong oscillatory behavior of nuclear form factors that becomes significant at higher momentum transfers, allowing a clean isolation of intrinsic nuclear-mass and form-factor effects~\cite{akimov2017,akimov2021,scholberg2006}.

The use of $\nu_e$ is a practical implementation choice within Geant4. For CEvNS, the interaction of interest is the neutral-current (NC) coherent elastic channel, which exists for all neutrino flavors and is largely flavor-independent at the level of recoil kinematics relevant for a relative target comparison. Consequently, the benchmark results presented here are intended to represent generic CEvNS recoil behavior rather than a source-specific flavor composition.

Only the neutral-current coherent elastic neutrino--nucleus scattering channel is included in the physics configuration. Charged-current (CC) channels are not considered because they do not correspond to coherent \emph{elastic} scattering: CC interactions change the nuclear charge and typically produce non-coherent or inelastic nuclear final states (and, depending on energy, additional charged leptons), which fall outside the CEvNS signal definition used in this work. Including CC processes would therefore mix fundamentally different interaction topologies with the CEvNS recoil sample and would not serve the purpose of a clean CEvNS benchmark comparison.

The neutrino beam is implemented using \texttt{G4ParticleGun}, with primaries generated along the $z$-axis and directed toward the target center to ensure uniform and reproducible illumination. For each target nucleus, $10^6$ primary neutrino events are simulated to achieve adequate statistical precision in the recoil-energy spectra and angular distributions.

Finally, no additional background sources (radiogenic neutrons, gamma backgrounds, cosmogenic secondaries) are simulated. This is an intentional scope choice: background modeling is highly detector- and site-dependent and is not required for the baseline comparison of intrinsic CEvNS recoil observables presented here. Background rejection, thresholds, and detector-response effects can be included in future, experiment-specific extensions, but are not necessary for the reference-level benchmarking objective of the present work.

\subsection{Physics Processes}

The standard Geant4 electromagnetic, decay, and optical physics lists were disabled except for minimal particle-transport requirements. 
A custom CEvNS interaction class (\texttt{CEvNSProcess}) was implemented to compute differential cross sections:

\[
\frac{d\sigma}{dT} = \frac{G_F^2}{4\pi}  Q_W^2  M 
\left(1 - \frac{M T}{2 E_\nu^2}\right) F^2(q) ,
\]

where $T$ is recoil energy, $M$ is nuclear mass, $Q_W$ is the weak charge, and $F(q)$ is the Helm form factor.  
Momentum-transfer dependence was treated through

\[
q = \sqrt{2 M T}.
\]

The Helm form factor was implemented as

\[
F(q) = 3\,\frac{j_1(qR_0)}{qR_0} e^{-\frac{(qs)^2}{2}},
\]

with $R_0$ the effective radius and $s$ the surface thickness.

\subsection{Data Recording}

A custom data manager recorded recoil energy, recoil position, scattering angle, momentum transfer, and event timing into ROOT files using TTrees. Fiducial and veto regions were tracked separately in order to maintain compatibility with realistic detector concepts; however, only fiducial recoils were considered in the present analysis, consistent with the detector-agnostic benchmark nature of this study.

\section{Detector Geometry}

The geometry adopted in this work is a deliberately simplified, reference-level configuration designed to support a controlled comparison of CEvNS recoil observables across different target nuclei. The central objective is to isolate intrinsic nuclear-physics effects—primarily target mass, neutron number, and nuclear form-factor suppression—under identical and reproducible simulation conditions, rather than to deliver an experiment-specific sensitivity or background-optimized detector design. For this reason, the simulated setup is used as a geometrical carrier that defines a well-posed target volume and a consistent beam--target configuration. Passive layers and auxiliary volumes (shielding, veto, and structural regions) are included only to define spatial boundaries and to allow unambiguous volume tagging of energy deposits, but they are not used to model background rejection or detector performance. In particular, background fields and sources (radiogenic neutrons, ambient gammas, cosmogenic secondaries) are not simulated because they are strongly site-, material-, and configuration-dependent and would require a dedicated, experiment-driven study beyond the scope of a comparative CEvNS benchmark. Similarly, detector-response effects—such as scintillation light yield, quenching, optical photon transport, photodetector response, trigger thresholds, and energy-resolution smearing—are intentionally omitted. Including such effects would introduce model and parameter dependence (e.g., light yield, collection efficiency, wavelength shifting, electronics response) that varies substantially across detector technologies and could mask the purely nuclear trends that are the focus of this work. Accordingly, the results reported here are extracted from the true nuclear-recoil information at the interaction level and should be interpreted as intrinsic CEvNS recoil distributions in a standardized geometry, suitable for benchmarking and material-comparison studies rather than for direct experimental rate predictions.

\section{Custom CEvNS Interaction Model}

The CEvNS interaction was implemented as a custom Geant4 discrete process (\texttt{CEvNSProcess}) derived from \texttt{G4VDiscreteProcess}. When an incident neutrino traverses the fiducial target volume, the process samples a nuclear recoil according to Standard-Model neutral-current CEvNS kinematics.

The recoil spectrum is generated from the standard differential cross section
\begin{equation}
\frac{d\sigma}{dT}(E_\nu,T)=\frac{G_F^2\,M}{4\pi}\,Q_W^2
\left(1-\frac{M T}{2E_\nu^2}\right)\,F^2(q),
\label{eq:cevns_diff}
\end{equation}
where $E_\nu$ is the incident neutrino energy, $T$ is the nuclear recoil energy, $M$ is the nuclear mass, and $q$ is the momentum transfer. The weak charge is taken as
\begin{equation}
Q_W = N - (1-4\sin^2\theta_W)\,Z,
\label{eq:weakcharge}
\end{equation}
and the maximum kinematically allowed recoil energy is
\begin{equation}
T_{\mathrm{max}}=\frac{2E_\nu^2}{M+2E_\nu}.
\label{eq:tmax}
\end{equation}

Loss of coherence at finite momentum transfer is included through the Helm nuclear form factor $F(q)$, and all nucleus-dependent quantities ($M$, $Z$, $N$, and $F(q)$) are computed separately for each target (B, Mg, Ti, Zr) to enable a direct comparison under identical simulation conditions. The implementation follows the standard CEvNS formalism used in the literature~\cite{freedman1974,scholberg2006,miranda2020,helm1956,lewin1996}.

\section{Results}

In this section, we present the nuclear recoil observables obtained from coherent elastic neutrino--nucleus scattering for the four candidate target nuclei: boron (B), magnesium (Mg), titanium (Ti), and zirconium (Zr). All simulations were performed under identical neutrino flux, geometry, and interaction conditions, with $10^{6}$ primary neutrino events generated for each target material. Nuclear recoil energies and scattering angles were recorded exclusively within the fiducial volume. By construction, detector-response effects and background contributions were excluded, ensuring that the observed differences originate solely from intrinsic CEvNS kinematics and nuclear structure effects. This controlled configuration allows a direct and unbiased comparison of target-dependent recoil behavior.

\subsection{Motivation for Target Nuclei Selection}

The selection of boron (B), magnesium (Mg), titanium (Ti), and zirconium (Zr) as candidate target nuclei is motivated by a combination of CEvNS kinematics, nuclear-structure effects, and material properties relevant for future low-threshold detector concepts. Rather than focusing on a single optimized target, this study aims to explore systematic trends across a representative range of nuclear masses and neutron numbers under controlled simulation conditions.

From a kinematic perspective, the maximum nuclear recoil energy in CEvNS scales inversely with the nuclear mass, favoring lighter nuclei for achieving higher recoil-energy endpoints. Boron and magnesium therefore serve as representative light and intermediate-mass targets that preserve sensitivity to higher recoil energies, which may be advantageous for detector technologies with finite energy thresholds. These nuclei provide access to recoil regimes where coherence is largely maintained and form-factor suppression remains minimal.

In contrast, heavier nuclei benefit from the approximate $N^2$ scaling of the CEvNS cross section in the coherent regime, leading to enhanced interaction rates at low recoil energies. Titanium and zirconium were selected as representative medium-to-heavy targets that balance increased neutron number with manageable coherence loss. Compared to very heavy nuclei, such as xenon or tungsten, Ti and Zr retain partial coherence over a broader recoil-energy range while still exhibiting pronounced form-factor effects. This makes them particularly suitable for studying the interplay between cross-section enhancement and nuclear-size suppression.

The chosen targets also span a wide range of nuclear radii and weak-charge distributions, allowing a systematic investigation of Helm form-factor effects across different momentum-transfer regimes. By including both light and heavier nuclei, the present study captures the transition from nearly fully coherent scattering to regimes where coherence loss becomes a dominant factor shaping recoil observables.

In addition to their nuclear-physics relevance, the selected elements are of practical interest for emerging detector technologies. Magnesium and titanium are commonly used structural and target materials, while zirconium-based compounds are widely employed in cryogenic and low-background applications due to their favorable mechanical and radiopurity properties. Boron-containing materials are also of interest for hybrid detector concepts and neutron-related applications. Although material-specific detector performance is not modeled in this benchmark study, these considerations motivate the selection of targets that are both physically informative and experimentally relevant.

Overall, the four selected nuclei form a complementary set that enables a controlled and systematic comparison of CEvNS recoil observables across a broad range of masses, neutron numbers, and form-factor behaviors. This approach provides a benchmark-level framework for evaluating target-material trade-offs in future CEvNS detector design studies, without bias toward a specific experimental implementation.

\subsection{CEvNS Recoil Energy Spectra}

Figure~X shows the nuclear recoil energy spectra obtained for the four target nuclei under identical simulation conditions. A clear mass-dependent trend is observed across the full recoil-energy range. Lighter nuclei such as boron and magnesium exhibit recoil spectra extending to higher maximum energies, while heavier targets such as titanium and zirconium produce spectra that are increasingly concentrated at lower recoil energies. This behavior follows directly from CEvNS kinematics, where the maximum recoil energy scales inversely with the nuclear mass, $T_{\mathrm{max}} \propto 1/M$.

In contrast to the kinematic suppression at high recoil energies, heavier nuclei benefit from an enhanced interaction probability at low recoil energies due to the approximate $N^2$ dependence of the CEvNS cross section in the coherent regime. As a result, titanium and zirconium exhibit larger event yields at low recoil energies compared to lighter targets, despite their reduced kinematic reach. This competition between kinematic effects and neutron-number scaling leads to distinct recoil-energy distributions for each nucleus.

At higher recoil energies, deviations from the simple $N^2$ scaling become increasingly pronounced, particularly for the heavier nuclei. This suppression arises from the loss of coherence at finite momentum transfer and is encoded through the nuclear form factor. The inclusion of the Helm form factor in the interaction model results in a progressive reduction of the differential cross section at large recoil energies, with the effect being strongest for zirconium and titanium due to their larger nuclear radii. Consequently, the recoil spectra of heavier targets fall off more rapidly at higher energies compared to those of boron and magnesium.

These results illustrate that light and heavy target nuclei offer complementary advantages for CEvNS detection. While heavy nuclei provide enhanced interaction rates at low recoil energies, lighter targets retain sensitivity to higher recoil energies that may be advantageous for detector technologies with finite energy thresholds. The recoil-energy spectra presented here therefore provide a quantitative benchmark for evaluating target-material trade-offs in future CEvNS detector studies.

\begin{figure}[H]
\centering

{\large \textbf{CEvNS Recoil Energy Spectra}}\\[0.5cm]

\begin{subfigure}{0.48\textwidth}
    \centering
    \includegraphics[width=\linewidth]{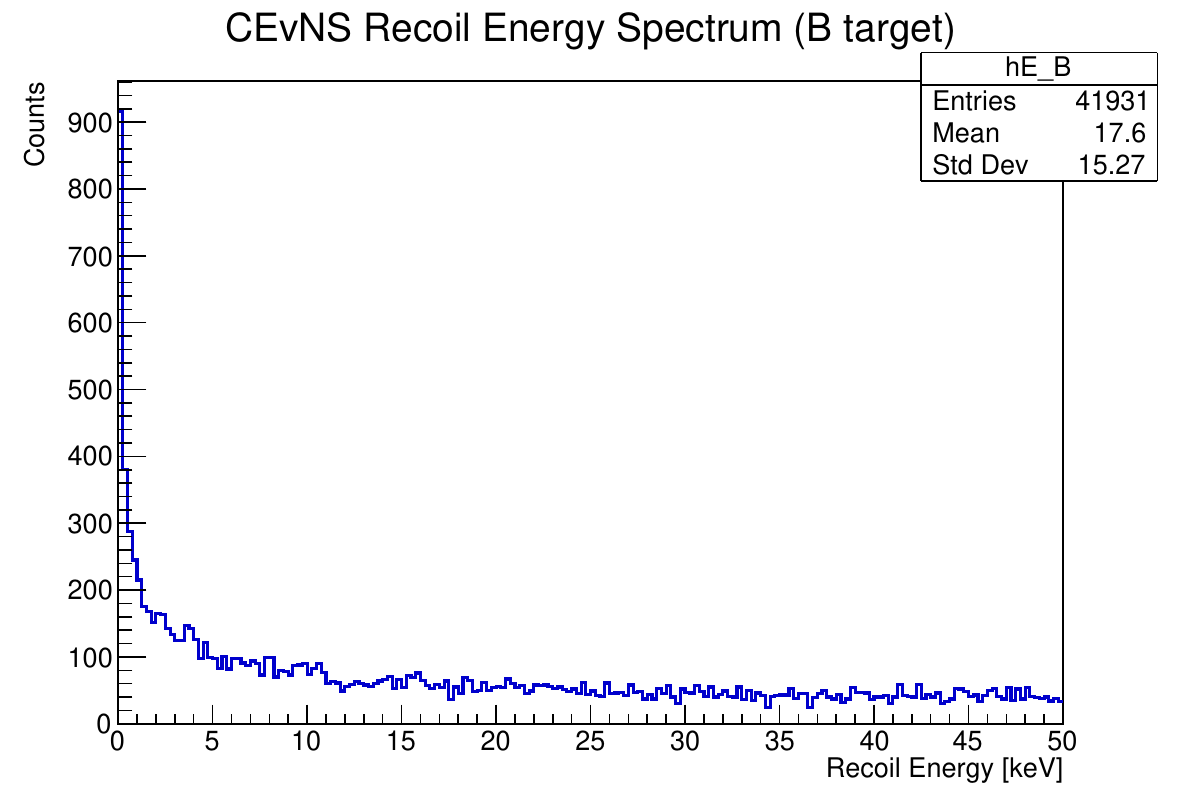}
    \caption{Boron (B) recoil energy spectrum.}
\end{subfigure}
\hfill
\begin{subfigure}{0.48\textwidth}
    \centering
    \includegraphics[width=\linewidth]{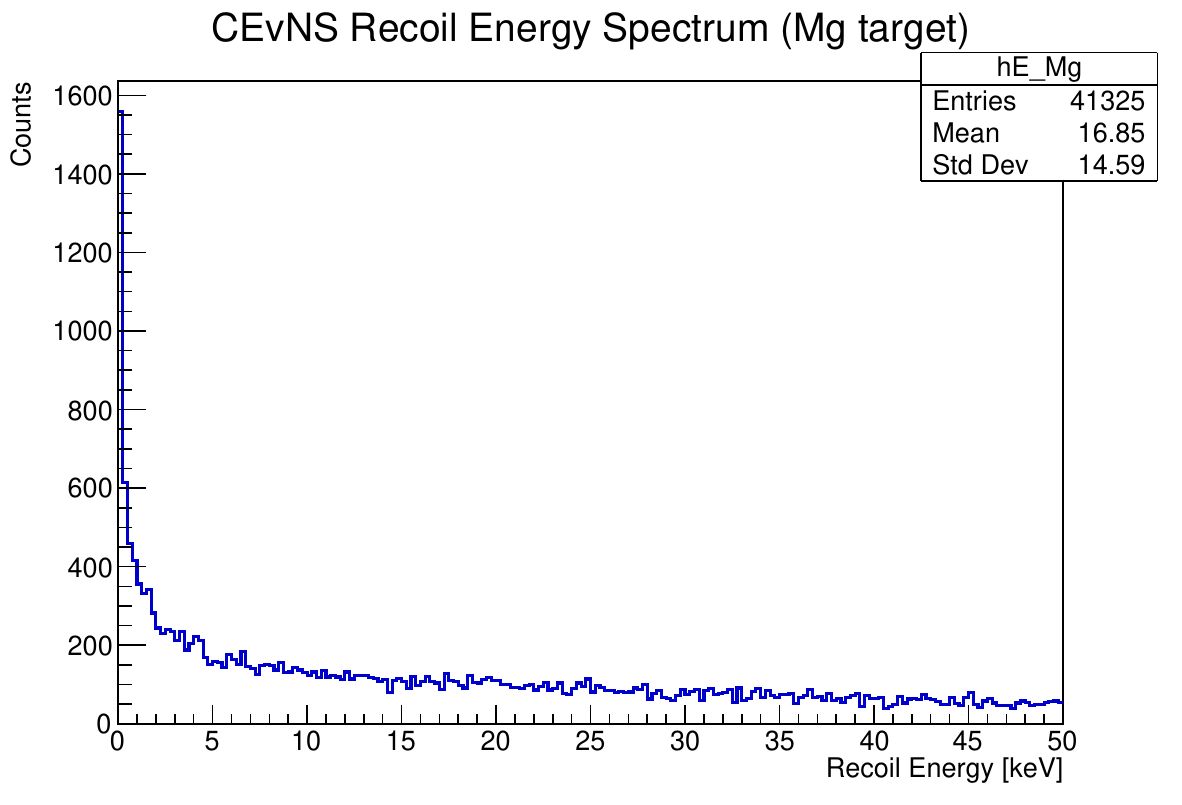}
    \caption{Magnesium (Mg) recoil energy spectrum.}
\end{subfigure}

\vspace{0.5cm}

\begin{subfigure}{0.48\textwidth}
    \centering
    \includegraphics[width=\linewidth]{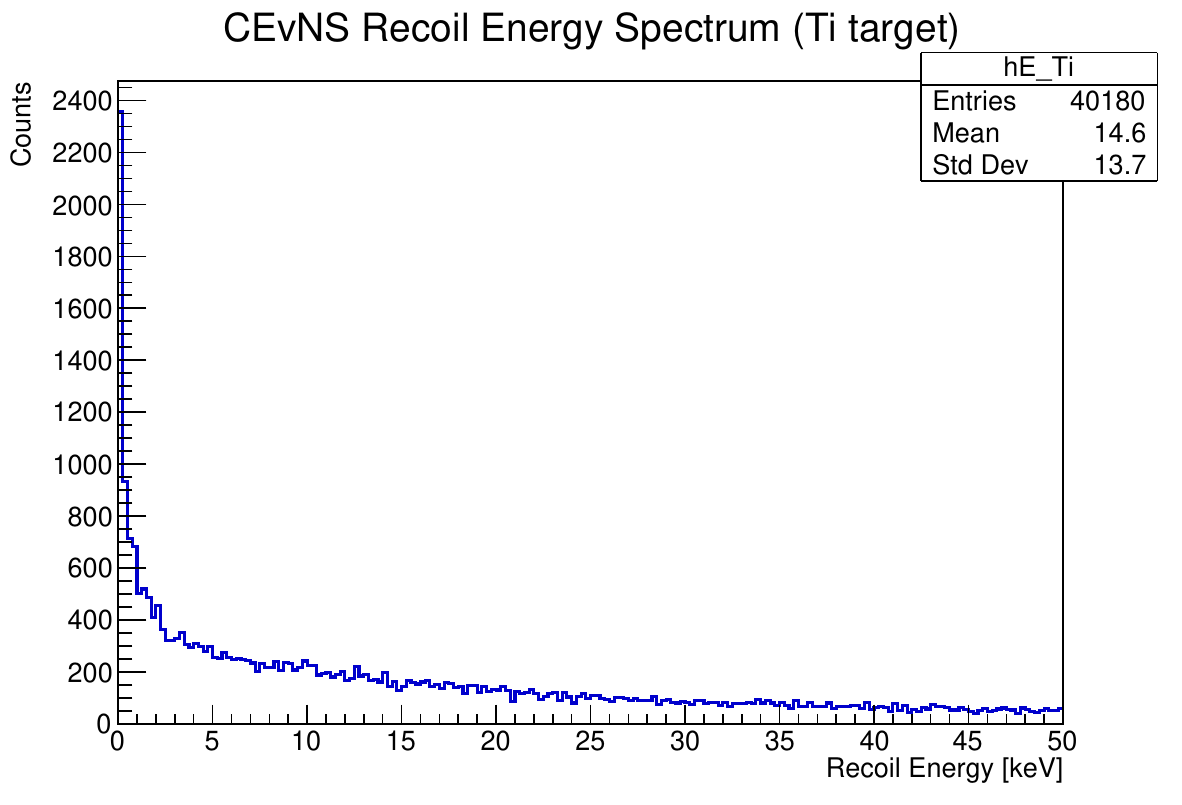}
    \caption{Titanium (Ti) recoil energy spectrum.}
\end{subfigure}
\hfill
\begin{subfigure}{0.48\textwidth}
    \centering
    \includegraphics[width=\linewidth]{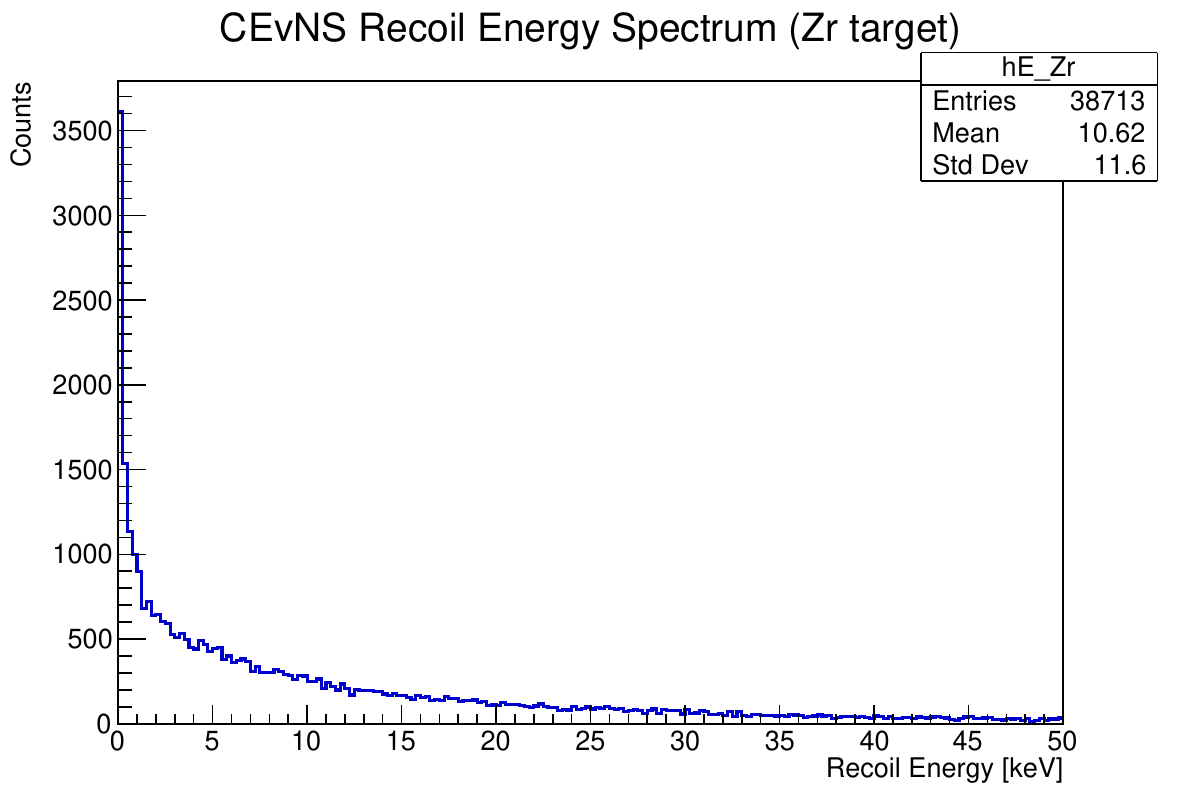}
    \caption{Zirconium (Zr) recoil energy spectrum.}
\end{subfigure}

\caption{Recoil energy spectra for the four proposed CEvNS target nuclei (B, Mg, Ti and Zr).}
\label{fig:recoil_energy}
\end{figure}

The recoil spectra in Fig.~\ref{fig:recoil_energy} display the expected mass dependence of CEvNS.
The light nuclei (B and Mg) produce harder spectra with higher maximum recoil energies and a
slowly falling tail, reflecting weaker form–factor suppression and more efficient transfer of
neutrino energy to the nucleus. In contrast, the heavier nuclei (Ti and especially Zr) exhibit
softer spectra that are strongly peaked at low recoil energies, since coherence is progressively
suppressed at larger momentum transfer. Under identical detector conditions, these trends imply
that light targets enhance recoil detectability, whereas heavy targets benefit from the larger
overall CEvNS event rate associated with their higher neutron numbers.

\subsection{Helm Form Factor Distributions}

The nuclear form factor plays a central role in CEvNS phenomenology by encoding the loss of coherence at finite momentum transfer. In this work, the Helm form factor is adopted as a physically motivated and computationally robust parametrization of the nuclear weak charge distribution.

The Helm model represents the nuclear density as a convolution of a hard uniform sphere with a Gaussian surface thickness, allowing it to capture the dominant features of finite nuclear size effects using a minimal set of parameters. This formulation provides a smooth and analytic description of coherence suppression, making it particularly suitable for Monte Carlo implementations where numerical stability and efficiency are essential.

Importantly, the Helm form factor has been shown to provide an accurate approximation to more detailed nuclear-structure models in the low to intermediate momentum-transfer regime relevant for CEvNS. In this kinematic domain, differences between commonly used form-factor parametrizations (such as symmetrized Fermi or Fourier–Bessel models) are subdominant compared to the leading nuclear-mass and recoil-energy dependencies. As a result, the Helm model is widely adopted in CEvNS studies as a benchmark choice that balances physical realism with model independence.

Given that the primary objective of this study is a controlled, nucleus-dependent comparison of recoil observables under identical simulation conditions, the Helm form factor provides a consistent and transparent baseline across all target nuclei considered. The use of a single, well-established parametrization ensures that observed differences in recoil spectra and angular distributions arise from intrinsic nuclear properties rather than from model-dependent variations in the assumed charge density.

The loss of coherence in coherent elastic neutrino--nucleus scattering at finite momentum transfer is quantified through the nuclear form factor, which accounts for the finite spatial extent of the nuclear weak-charge distribution. In this work, coherence effects are modeled using the Helm form factor, a commonly adopted parametrization that provides a smooth interpolation between the fully coherent regime at low momentum transfer and the incoherent regime at higher recoil energies.

\begin{figure}[H]
\centering

{\large \textbf{Helm Form Factor Distributions}}\\[0.5cm]

\begin{subfigure}{0.48\textwidth}
    \centering
    \includegraphics[width=\linewidth]{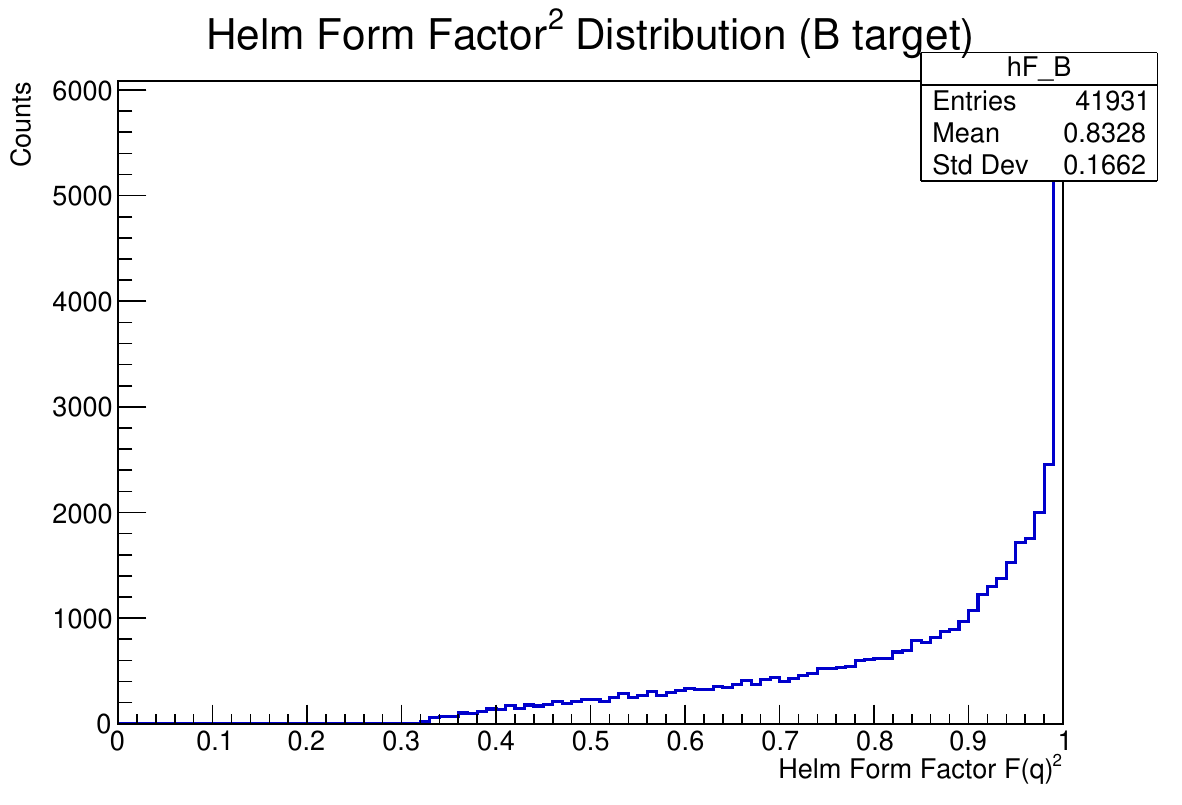}
    \caption{Boron (B)}
\end{subfigure}
\hfill
\begin{subfigure}{0.48\textwidth}
    \centering
    \includegraphics[width=\linewidth]{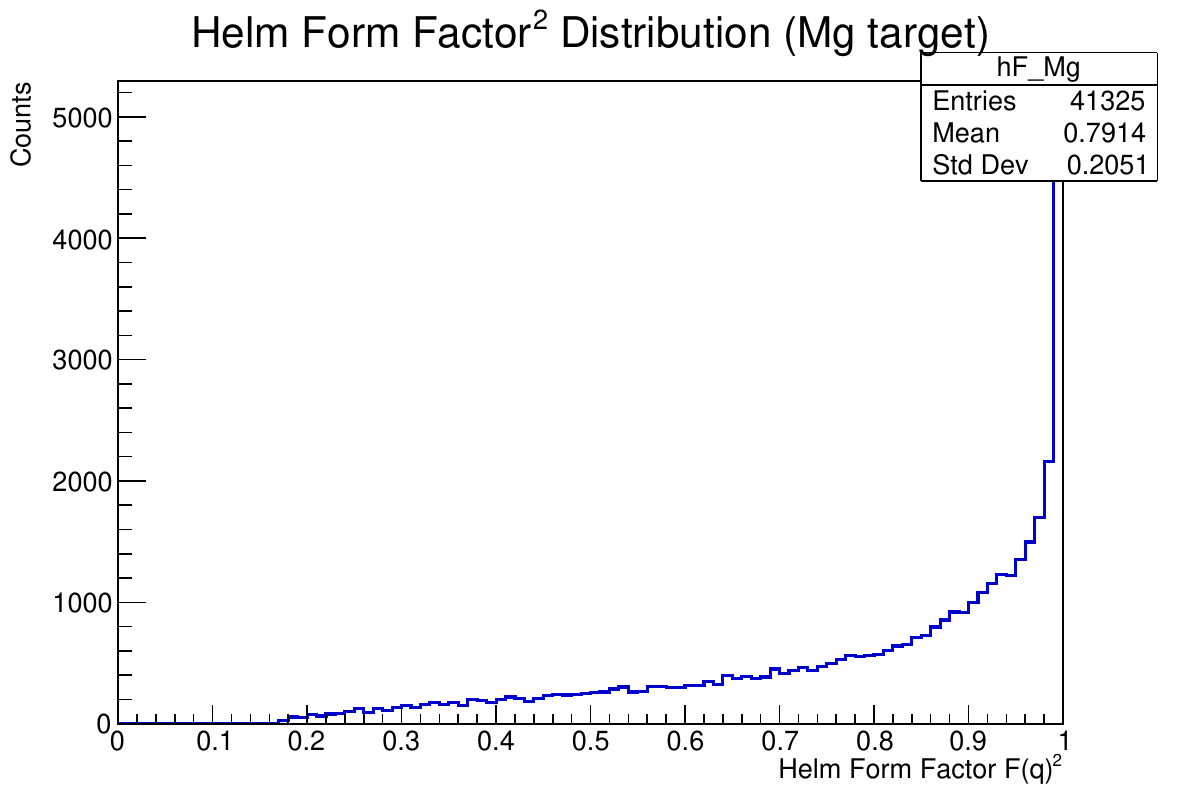}
    \caption{Magnesium (Mg)}
\end{subfigure}

\vspace{0.5cm}

\begin{subfigure}{0.48\textwidth}
    \centering
    \includegraphics[width=\linewidth]{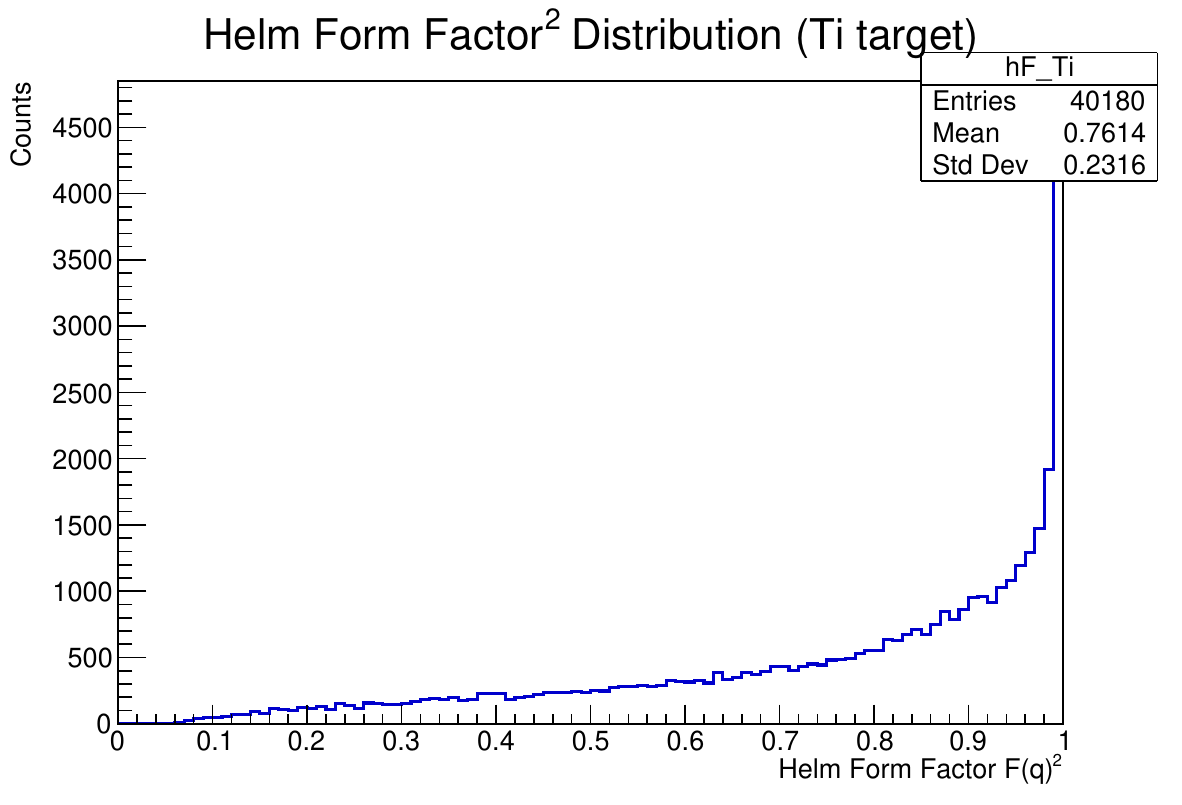}
    \caption{Titanium (Ti)}
\end{subfigure}
\hfill
\begin{subfigure}{0.48\textwidth}
    \centering
    \includegraphics[width=\linewidth]{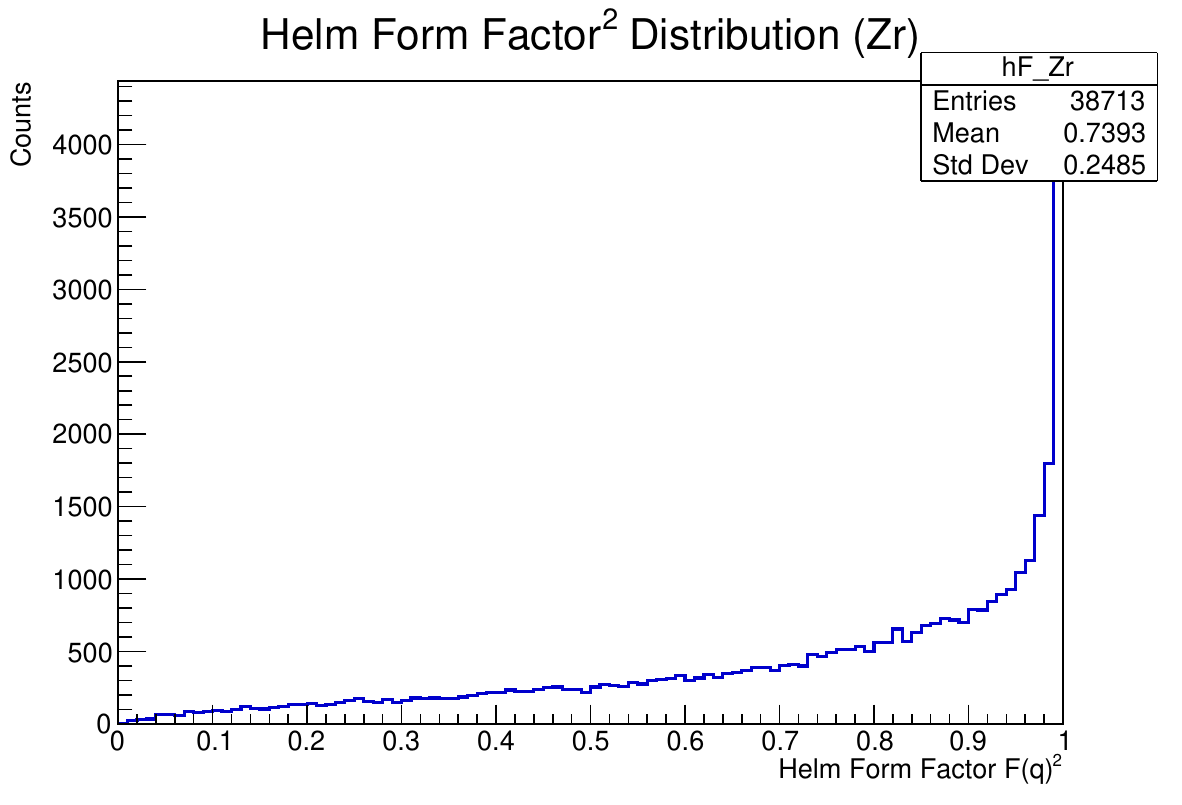}
    \caption{Zirconium (Zr)}
\end{subfigure}

\caption{Distributions of the squared Helm nuclear form factor, $F(q)^2$, for the four proposed CEvNS target nuclei (B, Mg, Ti, and Zr), obtained from Monte Carlo recoil kinematics. Lighter nuclei preserve coherence over a wider recoil-energy range, while heavier nuclei exhibit stronger form-factor suppression due to their larger nuclear radii.}
\label{fig:formfactor}
\end{figure}

Figure~\ref{fig:formfactor} shows the distributions of the squared Helm form factor, $F(q)^2$, obtained for the four target nuclei (B, Mg, Ti, and Zr) from Monte Carlo recoil kinematics. For light nuclei such as boron and magnesium, the momentum transfer associated with typical CEvNS recoils remains small compared to the inverse nuclear radius. As a consequence, coherence is largely preserved over the relevant recoil-energy range, resulting in $F(q)^2$ values close to unity for most events.

In contrast, heavier nuclei such as titanium and zirconium exhibit a significantly stronger suppression of $F(q)^2$. Their larger nuclear radii lead to a more rapid onset of coherence loss as the momentum transfer increases, causing a substantial reduction of the effective cross section at higher recoil energies. This behavior is reflected in the broader distributions toward lower $F(q)^2$ values observed for Ti and Zr.

These form-factor effects directly impact the recoil-energy spectra discussed in the previous section. While heavy nuclei benefit from an enhanced interaction rate at low recoil energies due to the approximate $N^2$ scaling of the CEvNS cross section, the suppression induced by the Helm form factor increasingly reduces their contribution at higher recoil energies. The combined influence of nuclear size and form-factor suppression therefore plays a central role in shaping the observed recoil spectra and must be carefully considered when assessing target-material performance in CEvNS detector studies.

\subsection{Nuclear Form Factor Suppression as a Function of Recoil Energy}

\begin{figure}[H]
\centering

{\large \textbf{Nuclear Form Factor $F(q)^2$ vs Recoil Energy}}\\[0.5cm]

\begin{subfigure}{0.48\textwidth}
    \centering
    \includegraphics[width=\linewidth]{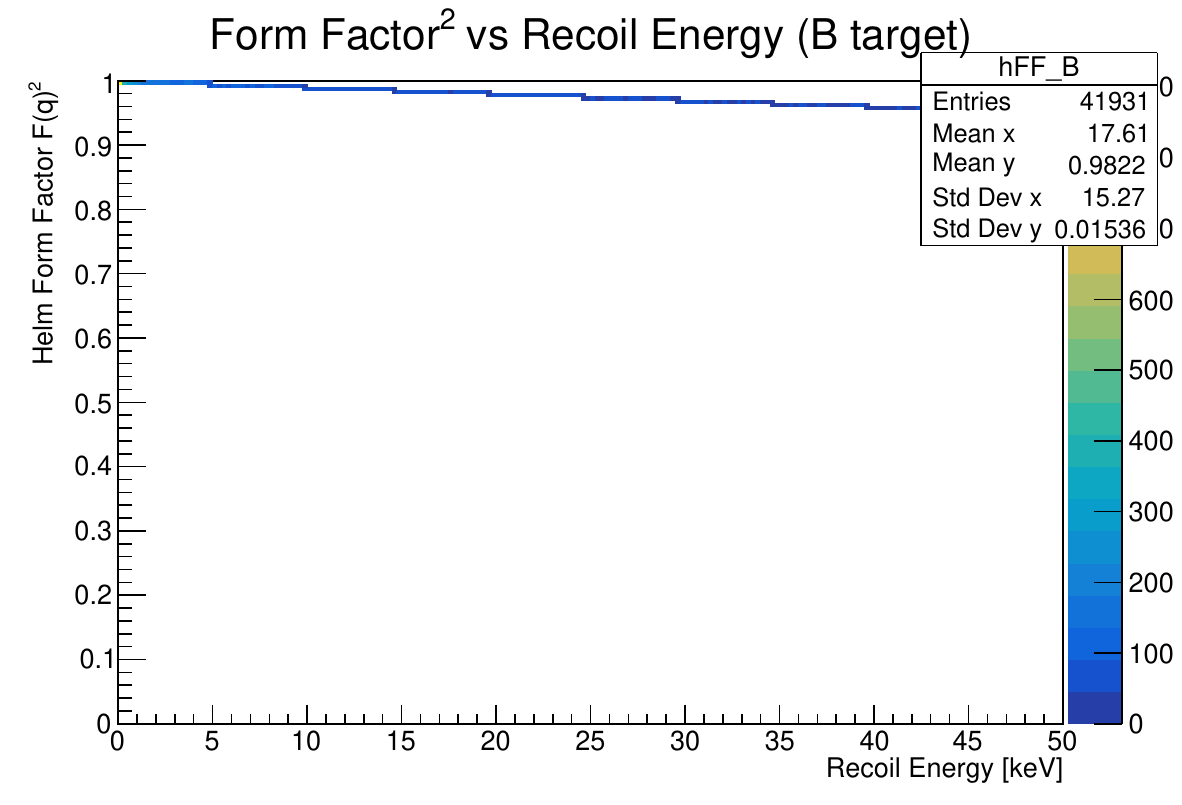}
    \caption{Boron (B) target}
\end{subfigure}
\hfill
\begin{subfigure}{0.48\textwidth}
    \centering
    \includegraphics[width=\linewidth]{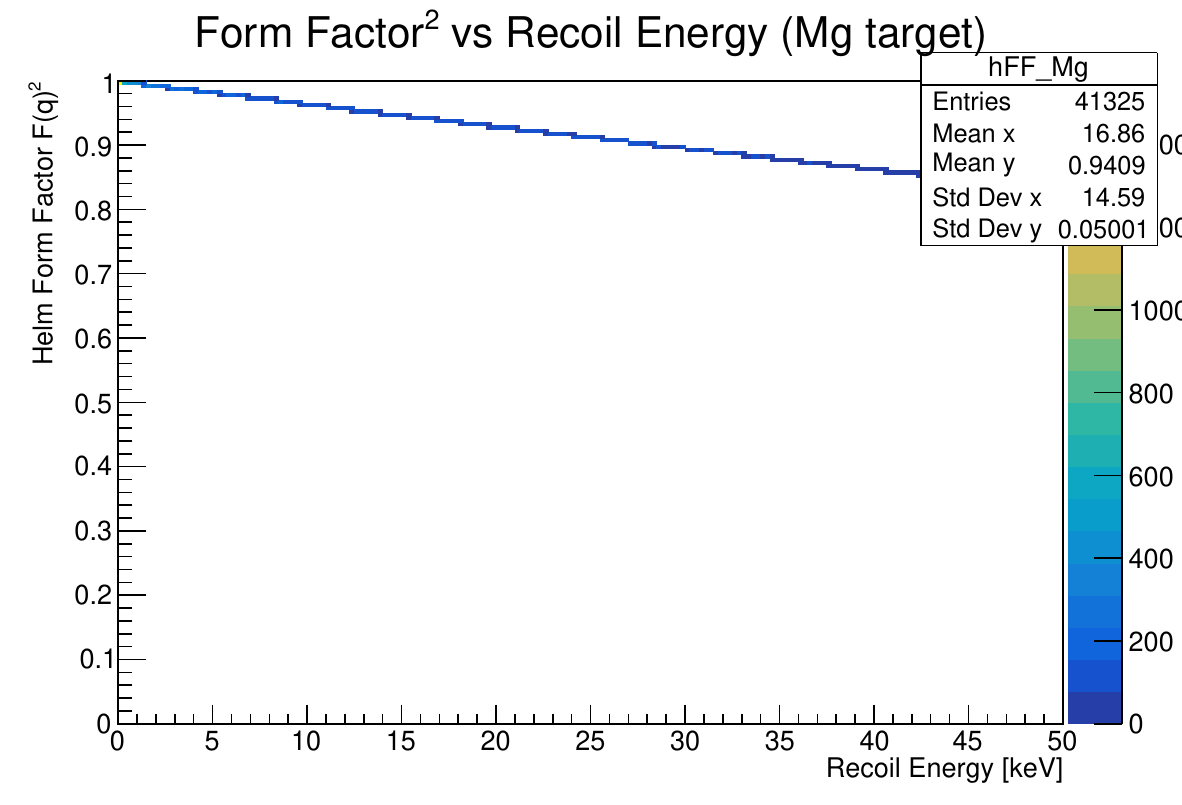}
    \caption{Magnesium (Mg) target}
\end{subfigure}

\vspace{0.5cm}

\begin{subfigure}{0.48\textwidth}
    \centering
    \includegraphics[width=\linewidth]{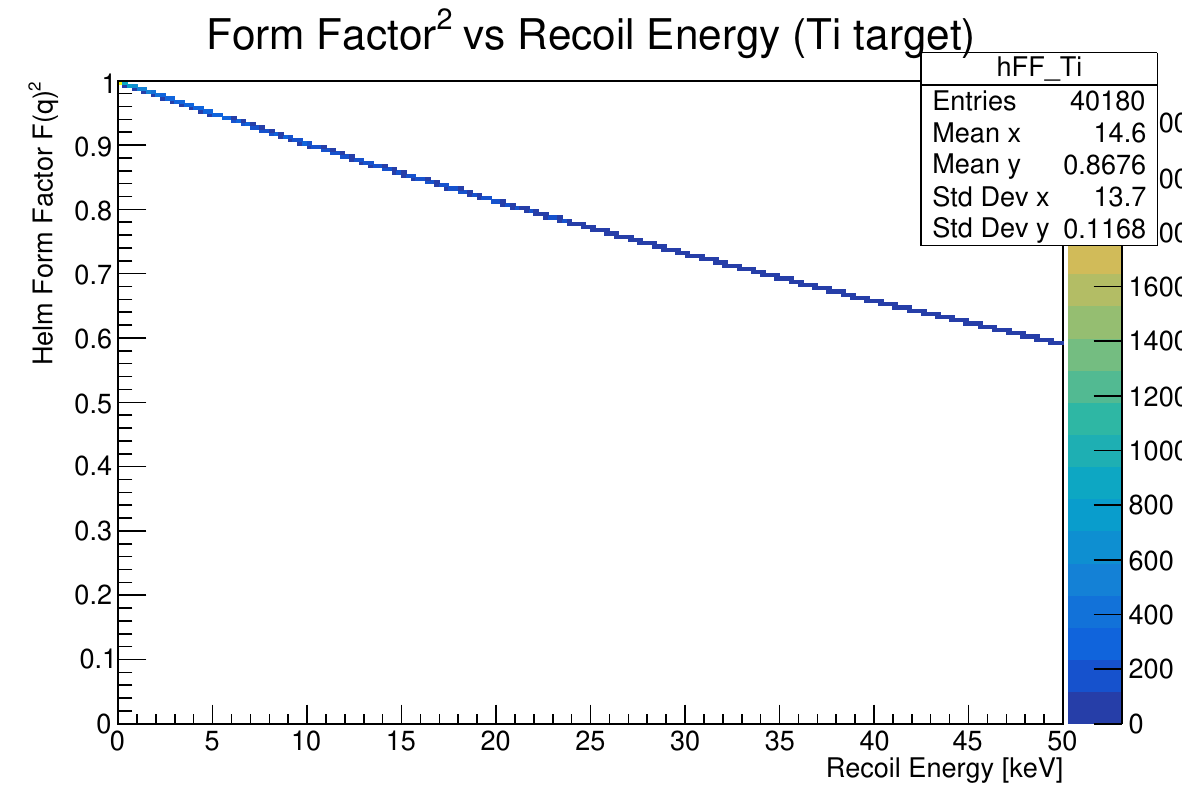}
    \caption{Titanium (Ti) target}
\end{subfigure}
\hfill
\begin{subfigure}{0.48\textwidth}
    \centering
    \includegraphics[width=\linewidth]{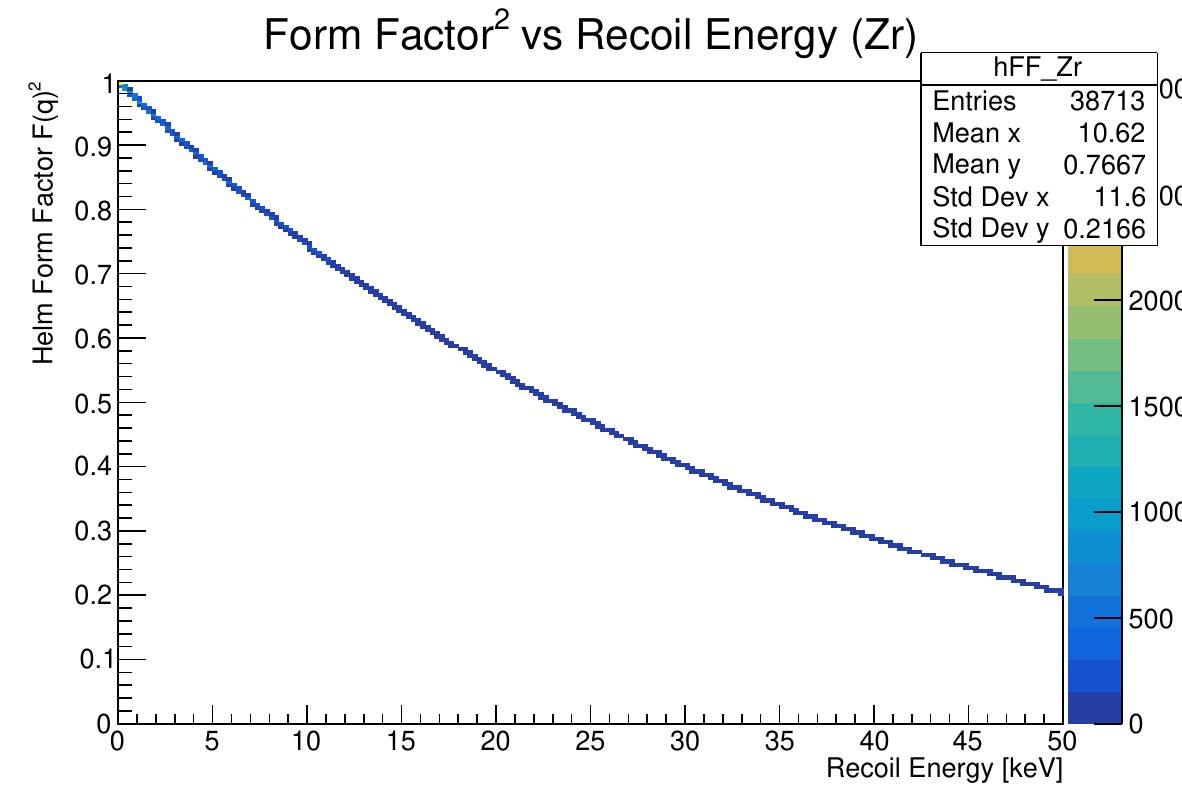}
    \caption{Zirconium (Zr) target}
\end{subfigure}

\caption{Squared Helm form factor $F(q)^2$ as a function of nuclear recoil energy
for the four proposed CEvNS target nuclei (B, Mg, Ti and Zr).}
\label{fig:formfactor_vs_energy}
\end{figure}

The distributions of the squared nuclear form factor $F(q)^2$ as a function of recoil
energy quantify how coherency is progressively lost as the momentum transfer increases.
Since the CEvNS differential cross section scales approximately as
\mbox{$\mathrm{d}\sigma/\mathrm{d}E_r \propto N^2\,F(q)^2$}, any deviation of $F(q)^2$
from unity translates directly into a suppression of the observable recoil rate.
For the light nuclei, boron and magnesium, the Helm form factor remains close to
$F(q)^2 \simeq 1$ over most of the $0$--$50$~keV recoil window.  In this regime the
scattering is almost perfectly coherent, and the recoil spectra for B and Mg are
dominated by neutrino kinematics with only a modest nuclear-structure correction.

For titanium and especially zirconium, the behaviour is markedly different and reflects
the larger nuclear size.  At the same recoil energy the corresponding momentum transfer
is comparable to, or larger than, the inverse nuclear radius, and the oscillatory
structure of the Helm form factor leads to a substantial reduction of $F(q)^2$.
In the Zr case, a significant fraction of events at ${\cal O}(10\,\text{keV})$ already
experience $F(q)^2<1$, indicating that part of the naive $N^2$ enhancement is ``eaten
up'' by coherence loss.  Titanium shows an intermediate trend, with the onset of
suppression occurring earlier than for Mg but remaining less severe than for Zr.

Taken together, these results highlight the competition between neutron-number
enhancement and form-factor suppression in CEvNS.  Light targets (B, Mg) maintain nearly
full coherence across a broad recoil-energy range and are therefore well suited for
experiments aiming at spectral measurements extending to higher energies.  Heavy targets
(Ti, Zr) retain their advantage only at very low recoil energies where $F(q)^2 \approx 1$,
and their sensitivity degrades rapidly once the recoil energy enters the region where
form-factor effects dominate.  This behaviour is crucial for assessing which nuclei are
optimal for future low-threshold CEvNS detector concepts.

\subsection{Recoil Angular Distributions}

In addition to recoil-energy observables, the angular distributions of CEvNS-induced nuclear recoils were examined in order to validate the kinematic consistency of the implemented interaction model across different nuclear targets. For coherent elastic neutrino--nucleus scattering, the nuclear recoil directions are expected to be strongly forward-peaked in the laboratory frame, reflecting the small momentum transfer characteristic of low-energy neutrino interactions.

\begin{figure}[H]
\centering
{\large \textbf{CEvNS Scattering-Angle Distributions}}\\[0.5cm]

\begin{subfigure}{0.48\textwidth}
    \centering
    \includegraphics[width=\linewidth]{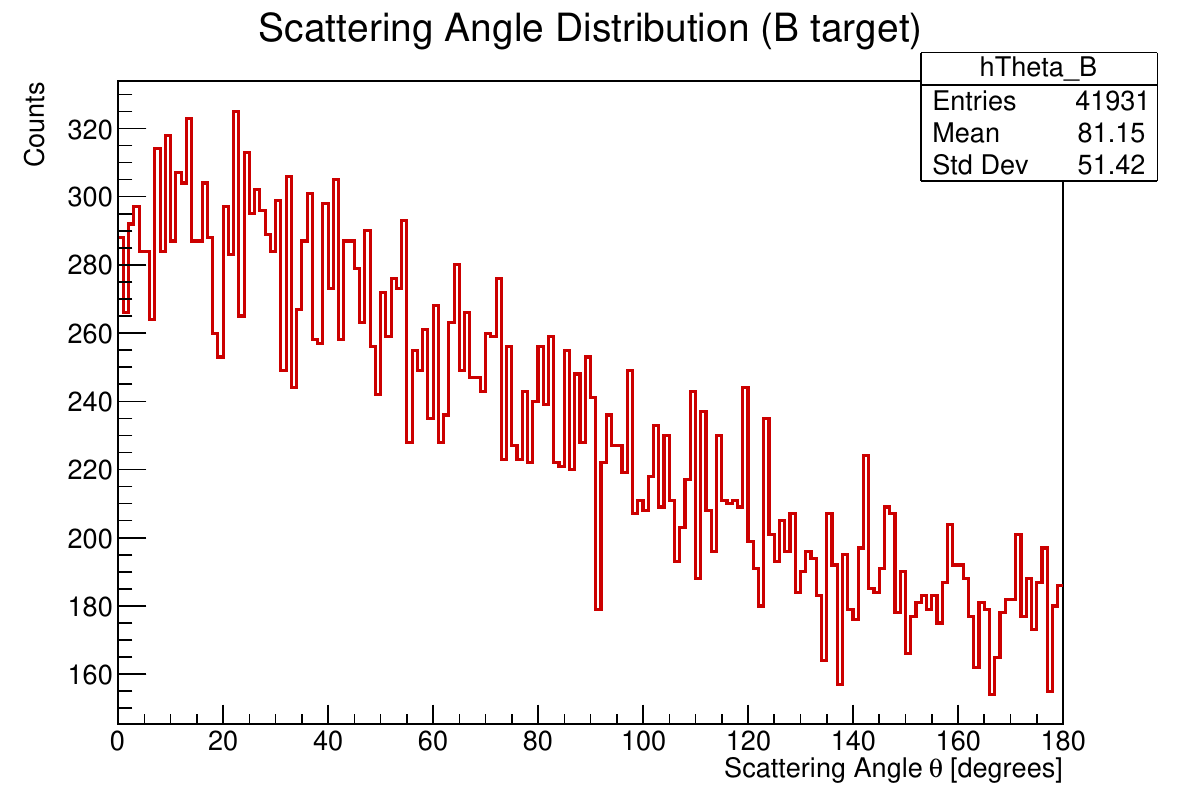}
    \caption{Boron}
\end{subfigure}
\hfill
\begin{subfigure}{0.48\textwidth}
    \centering
    \includegraphics[width=\linewidth]{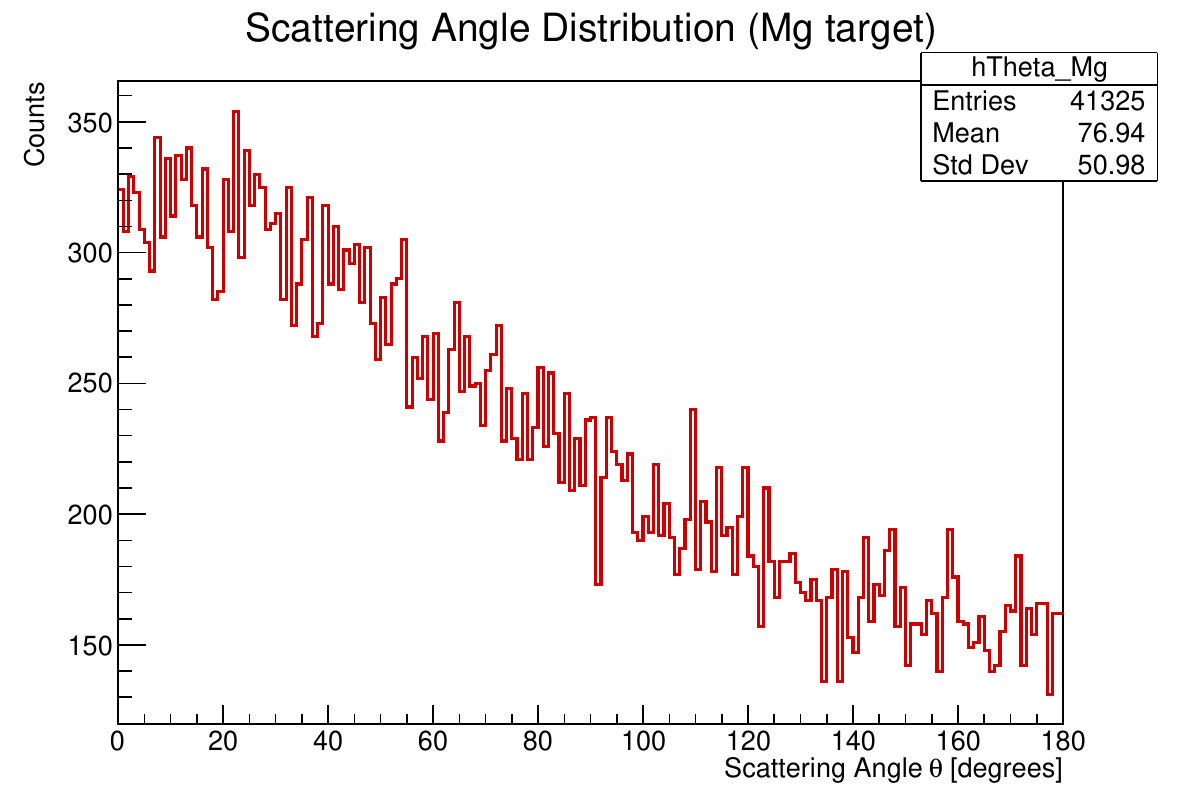}
    \caption{Magnesium}
\end{subfigure}

\vspace{0.5cm}

\begin{subfigure}{0.48\textwidth}
    \centering
    \includegraphics[width=\linewidth]{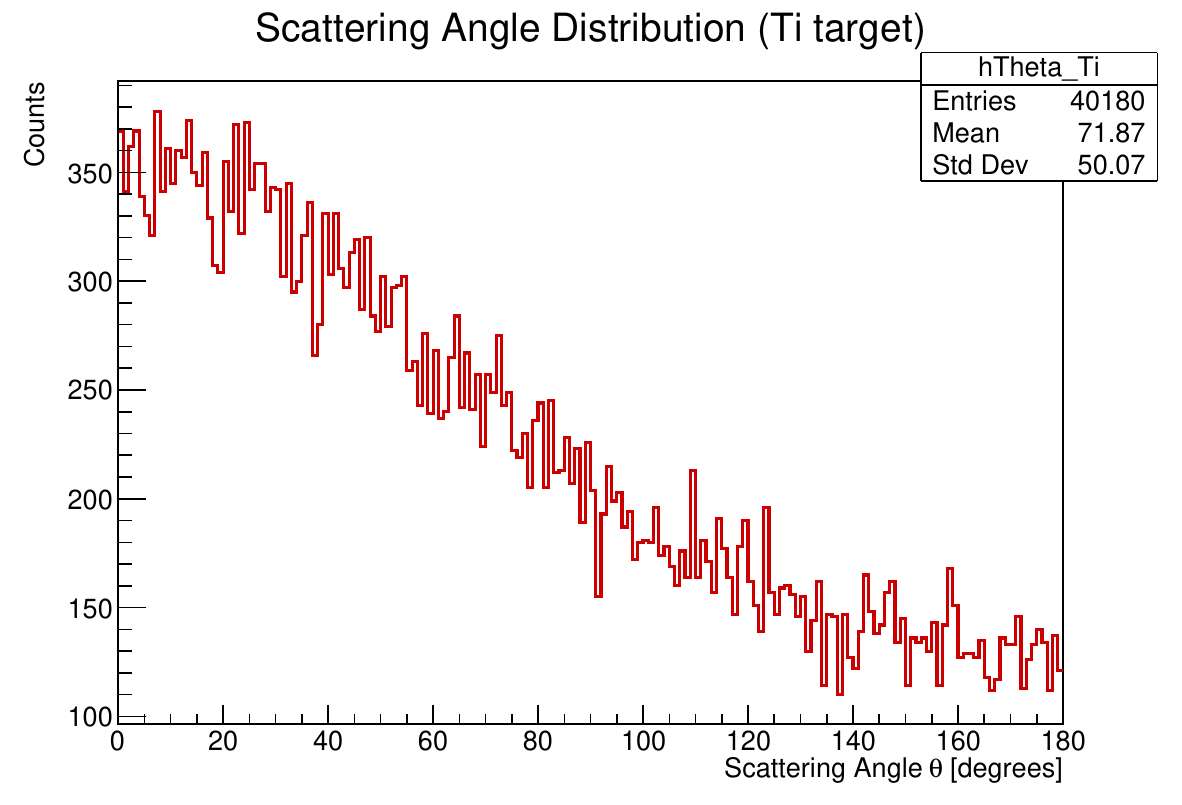}
    \caption{Titanium}
\end{subfigure}
\hfill
\begin{subfigure}{0.48\textwidth}
    \centering
    \includegraphics[width=\linewidth]{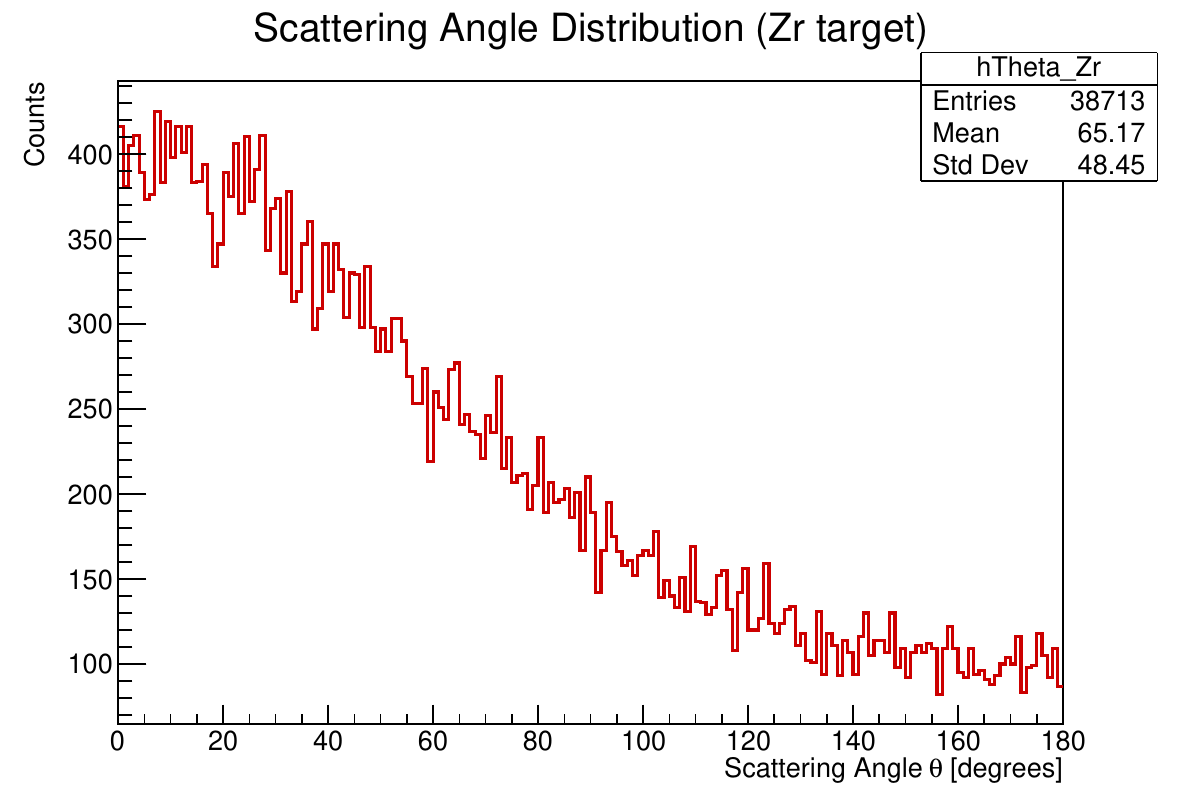}
    \caption{Zirconium}
\end{subfigure}

\caption{Scattering–angle distributions of CEvNS–induced nuclear recoils for the four
proposed target nuclei (B, Mg, Ti, and Zr). Each panel shows the event counts as a
function of polar scattering angle $\theta$ in the laboratory frame, illustrating the
forward–peaked but broadly distributed nature of CEvNS recoils in these materials.}
\label{fig:theta_distributions}
\end{figure}

As shown in Fig.~\ref{fig:theta_distributions}, all four target nuclei (B, Mg, Ti, and Zr) exhibit forward-biased angular distributions, consistent with two-body CEvNS kinematics. Lighter nuclei display slightly broader angular spreads, which arise from their larger recoil energies and wider kinematically allowed phase space. In contrast, heavier nuclei show a more pronounced forward collimation due to their larger nuclear masses, which suppress large-angle recoils.

While angular information is not exploited for event selection or background discrimination in the present benchmark study, these distributions serve an important validation role. In particular, the observed target-dependent angular trends confirm that the momentum-transfer dependence and recoil kinematics are being treated consistently within the custom CEvNS interaction model. The inclusion of angular observables therefore provides an additional cross-check of the physical correctness of the simulation framework, independent of detector-specific response effects.

We emphasize that the present analysis intentionally decouples intrinsic CEvNS kinematics from experiment-specific angular reconstruction capabilities. Consequently, angular distributions are reported here as a physics-level benchmark rather than as optimized detector observables. Their relevance for veto design, directional sensitivity, and background rejection in realistic detector geometries is left for future, experiment-driven studies.

\section{Discussion}

The purpose of this study is not to propose an optimized or experiment-specific CEvNS detector design, but to establish a nucleus-dependent benchmark of intrinsic CEvNS recoil observables under fully controlled and identical simulation conditions. By deliberately decoupling detector-response effects from nuclear-physics--driven behavior, this framework is intended to serve as a reusable reference for future detector-specific studies rather than a finalized experimental proposal. This benchmark-driven philosophy follows the original formulation of coherent elastic neutrino--nucleus scattering, where the dominant physics is governed by nuclear mass, neutron number, and form-factor effects rather than detector technology itself \cite{freedman1974,scholberg2006}.

The results presented in this work provide a systematic and detector-agnostic comparison of CEvNS recoil observables across four target nuclei spanning a broad range of nuclear masses and neutron numbers. Rather than optimizing a specific detector configuration, the analysis isolates the impact of intrinsic nuclear properties on recoil-energy spectra, angular distributions, and form-factor suppression under identical simulation conditions. This benchmark-level approach enables a clear interpretation of how target-dependent effects shape CEvNS observables independently of detector-response assumptions, complementing previous CEvNS studies that focus on individual materials or experiment-specific detector geometries \cite{akimov2017,akimov2021}.

A central outcome of the study is the identification of complementary roles played by light and heavier nuclei in CEvNS detection. Light targets such as boron and magnesium are characterized by higher recoil-energy endpoints, which arise directly from CEvNS kinematics and the inverse dependence of the maximum recoil energy on nuclear mass \cite{freedman1974}. These features make lighter nuclei particularly attractive for detector concepts that rely on large recoil signals or operate with limited energy resolution. In addition, the relatively small nuclear radii of light targets ensure that coherence is preserved over a wide momentum-transfer range, resulting in minimal Helm form-factor suppression across most of the recoil spectrum \cite{helm1956}.

Heavier nuclei, represented here by titanium and zirconium, exhibit contrasting behavior driven primarily by their larger neutron numbers. The approximate $N^{2}$ scaling of the CEvNS cross section enhances interaction rates at low recoil energies, leading to substantially higher event statistics despite reduced kinematic reach \cite{scholberg2006}. Although form-factor suppression becomes increasingly relevant for heavier nuclei at higher recoil energies, the present results demonstrate that coherence loss remains moderate within the energy range typical of reactor and stopped-pion neutrino sources. This balance between enhanced rates and manageable coherence loss highlights the suitability of medium-to-heavy nuclei for CEvNS detector concepts designed to maximize event counts.

The interplay between kinematic effects and nuclear-structure suppression underscores the importance of target selection as a detector-design parameter rather than a purely material choice. No single nucleus emerges as universally optimal; instead, the preferred target depends on the relative emphasis placed on recoil-energy visibility versus interaction statistics. The results therefore motivate hybrid or multi-target detection strategies in which complementary nuclei are employed to probe different regions of the CEvNS phase space.

Regarding novelty and scope, while individual aspects of CEvNS target dependence have been explored in previous works, the novelty of the present study lies in its systematic and detector-agnostic benchmarking of multiple candidate target nuclei under fully identical and controlled simulation conditions. By deliberately decoupling detector-response effects from intrinsic nuclear physics, this work provides a unified framework in which recoil-energy spectra, angular distributions, and form-factor suppression effects can be compared on equal footing across a wide nuclear-mass range. Rather than proposing an optimized detector design for a specific experimental configuration, the study establishes a set of physically transparent selection criteria that clarify how different nuclear targets populate complementary regions of the CEvNS phase space.

In this sense, the results serve as a material-selection guideline for future low-threshold CEvNS detectors, enabling informed target choices based on the relative prioritization of recoil-energy visibility, interaction statistics, and coherence preservation. This benchmark-level perspective is intended to complement, rather than replace, experiment-specific optimization studies, and provides a reusable reference for the design and interpretation of next-generation CEvNS experiments.

Finally, it is important to emphasize that the present conclusions are intentionally decoupled from detector-specific performance metrics. Effects such as energy thresholds, quenching, optical response, and background rejection were excluded by design in order to focus on intrinsic CEvNS physics. As such, the trends identified here should be interpreted as guiding principles for target-material selection, to be combined with detailed detector modeling in future experiment-specific studies. Overall, this work establishes a physically transparent benchmark for comparing CEvNS target nuclei and clarifies the trade-offs governing recoil observables across different nuclear regimes, providing a foundation for informed target selection in next-generation low-threshold CEvNS experiments.

\section{Conclusion}

We presented a benchmark-level Monte Carlo study of coherent elastic neutrino--nucleus scattering (CEvNS) recoil observables for four candidate target nuclei: boron (B), magnesium (Mg), titanium (Ti), and zirconium (Zr). A unified Geant4--ROOT simulation framework was employed to ensure a controlled and systematic comparison under identical interaction geometry and analysis selections. The CEvNS interaction was implemented with two-body recoil kinematics and nuclear-structure effects were incorporated through the Helm form factor, enabling a consistent treatment of coherence loss at finite momentum transfer \cite{freedman1974,lewin1996}.

The results highlight the expected and practically relevant trade-offs between light and heavy nuclei. Light targets (B, Mg) exhibit higher recoil-energy endpoints, reflecting the inverse dependence of the maximum recoil energy on nuclear mass, and preserve coherence over a broader fraction of the recoil phase space due to their smaller nuclear radii \cite{scholberg2006,akimov2021}. These features make light nuclei favorable for detector concepts emphasizing recoil-energy visibility and operation near threshold.

In contrast, heavier nuclei (Ti, Zr) benefit from the approximate $N^{2}$ scaling of the CEvNS cross section, leading to enhanced interaction rates at low recoil energies despite reduced kinematic reach \cite{freedman1974,akimov2017}. Although form-factor suppression becomes increasingly relevant for heavier nuclei at higher momentum transfer, the present results demonstrate that coherence loss remains moderate within the recoil-energy ranges characteristic of reactor and stopped-pion neutrino sources \cite{akimov2021}. This balance between increased event statistics and manageable coherence suppression highlights the suitability of medium-to-heavy nuclei for compact CEvNS detectors designed to maximize interaction rates.

The interplay between recoil kinematics and nuclear-structure effects underscores the importance of target selection as a detector-design parameter rather than a purely material choice. No single nucleus emerges as universally optimal; instead, the preferred target depends on the relative emphasis placed on recoil-energy visibility versus interaction statistics. The results therefore motivate hybrid or multi-target strategies in which complementary nuclei are employed to probe different regions of CEvNS phase space \cite{scholberg2006}.

Overall, this framework establishes a physically transparent and detector-agnostic benchmark for comparing CEvNS target nuclei. By intentionally decoupling detector-response effects such as energy thresholds, quenching, optical response, and background rejection, the present study isolates intrinsic nuclear-physics trends. As such, the conclusions should be interpreted as guiding principles for target-material selection, to be combined in future work with detailed, experiment-specific detector modeling \cite{akimov2017}.

\section*{Acknowledgements}

The author acknowledges the Geant4 Collaboration and the ROOT Development Team for providing the simulation and data-analysis frameworks used in this study. 
The scientific content, including the methodology, results, and interpretations, was developed entirely by the author. 
ChatGPT was used solely for language refinement and clarity enhancement and did not contribute to the scientific content of the manuscript.

\bibliographystyle{unsrt}
\bibliography{cevs_refs}

\end{document}